\documentclass[12pt]{iopart}

\usepackage{graphicx}
\usepackage{tikz}
\usepackage{iopams}  

\def\be{\begin{equation}}
\def\ee{\end{equation}}
\def\ba{\begin{eqnarray}}
\def\ea{\end{eqnarray}}

\begin{document}

\title[Critical Potts domain walls]{Bulk and boundary critical
  behaviour of thin and thick domain walls in the two-dimensional
  Potts model}

\author{J\'er\^ome Dubail$^{1,2}$, Jesper Lykke Jacobsen$^{2,3}$ and Hubert Saleur$^{1,4}$}
\address{${}^1$Institut de Physique Th\'eorique, CEA Saclay,
91191 Gif Sur Yvette, France}
\address{${}^2$LPTENS, \'Ecole Normale Sup\'erieure, 24 rue Lhomond, 75231 Paris, France}
\address{${}^3$Universit\'e Pierre et Marie Curie, 4 place Jussieu, 75252 Paris, France}
\address{${}^4$Department of Physics,
University of Southern California, Los Angeles, CA 90089-0484}

\eads{\mailto{jerome.dubail@cea.fr}, 
      \mailto{jesper.jacobsen@ens.fr},
      \mailto{hubert.saleur@cea.fr}}

\begin{abstract}

 The geometrical critical behaviour of the two-dimensional $Q$-state
 Potts model is usually studied in terms of the Fortuin-Kasteleyn (FK)
 clusters, or their surrounding loops.  In this paper we study a quite
 different geometrical object: the spin clusters, defined as connected
 domains where the spin takes a constant value.  Unlike the usual
 loops, the domain walls separating different spin clusters can cross
 and branch. Moreover, they come in two versions, `thin' or 'thick',
 depending on whether they separate spin clusters of different or
 identical colours. For these reasons their critical behaviour is
 different from, and richer than, those of FK clusters.  We develop a
 transfer matrix technique enabling the formulation and numerical
 study of spin clusters even when $Q$ is not an integer. We further
 identify geometrically the crossing events which give rise to
 conformal correlation functions. We study the critical behaviour both
 in the bulk, and at a boundary with free, fixed, or mixed boundary
 conditions. This leads to infinite series of fundamental critical
 exponents, $h_{\ell_1-\ell_2,2\ell_1}$ in the bulk and
 $h_{1+2(\ell_1-\ell_2),1+4 \ell_1}$ at the boundary, valid for $0 \le
 Q \le 4$, that describe the insertion of $\ell_1$ thin and $\ell_2$
 thick domain walls. We argue that these exponents imply that the
 domain walls are `thin' and `thick' also in the continuum limit.
 A special case of the bulk exponents is derived analytically from
 a massless scattering approach.

\end{abstract}


\section{Introduction}

Many of the key developments in the study of two-dimensional critical
phenomena originate from the study of only a very few lattice models
\cite{Wu08,Baxter82}. These include dimer coverings, the Ising and
six-vertex models, the O($n$) model, and the $Q$-state Potts model.
The latter two models are particularly important, as they can be
formulated in terms of geometrical degrees of freedom that in turn
describe extended fluctuating objects, such as domain walls in
magnets, percolation clusters, and polymers adsorbed on walls and
interfaces \cite{Eisenriegler}. These objects are in fact conformally
invariant whenever $n$ or $\sqrt{Q}$ is comprised in the range $-2 < n
\le 2$, and their fluctuations are characterised by critical
exponents.

The purpose of this paper is to revisit the geometrical formulation of
the two-dimensional Potts model. In particular, we shall define a set
of geometrical observables which are different from those considered
in most of the existing literature. These observables are simply the
domain walls in the formulation of the Potts model in terms of
$Q$-component spins. Not only are they in many ways more natural and
physically more relevant than the Fortuin-Kasteleyn (FK) clusters
\cite{FK72} usually considered, but they also turn out to have a richer
critical behaviour and a more complete set of critical exponents.
These features stem in part from the fact that there are actually two
different types of spin domain walls (thin and thick), as we shall
explain shortly.

The $Q$-state Potts model is defined by the partition function
\be
 Z = \sum_{\sigma} \prod_{(ij) \in E}
 \exp \left( K \delta_{\sigma_i,\sigma_j} \right) \,,
 \label{Potts}
\ee
where $K$ is the coupling between spins $\sigma_i = 1,2,\ldots,Q$
along the edges $E$ of some lattice ${\cal L}$. For simplicity we
shall take ${\cal L}$ to be the square lattice in the computations
below, whereas we use the triangular lattice in the figures. The
Kronecker delta function $\delta_{\sigma_i,\sigma_j}$ equals 1 if
$\sigma_i = \sigma_j$, and 0 otherwise.

The usual route \cite{FK72} is to write the obvious identity
\be
 \exp \left( K \delta_{\sigma_i,\sigma_j} \right) =
 1 + v \delta_{\sigma_i,\sigma_j} \,,
\ee
with $v = {\rm e}^K - 1$, and to expand $Z$ in powers of $v$.  The
result is an expression of $Z$ as a sum over FK clusters with weight
$v$ per unit length and fugacity $Q$ per connected
component. Alternatively one can think of the FK clusters in terms of
their surrounding hulls, which are self and mutually avoiding loops
with fugacity $n=\sqrt{Q}$ \cite{Baxter82}. Most features of these FK
clusters and loops, in the critical regime $-2 < n \le 2$, are by now
under complete control, thanks to the combined powers of Conformal
Field Theory (CFT) and Schramm-Loewner Evolution (SLE)
\cite{JJreview,BBreview}.  In particular, the loops behave like the
SLE trace in the continuum limit \cite{BBreview}, and viewing them as
contour lines of a (deformed) Gaussian free field leads to the Coulomb
Gas (CG) approach to CFT \cite{JJreview,Nienhuis}. Our understanding
of the critical properties of FK clusters and loops can be considered
almost complete, although some of their more intricate
observables---such as the backbone \cite{backbone} and shortest-path
\cite{shortest_path} dimensions---are still unknown.

The expansion in powers of $v$ has some pleasing features, notably
that the FK clusters coincide with percolation clusters in the formal
limit $Q \to 1$. But apart from that it is somewhat artificial in view
of the original formulation (\ref{Potts}) in terms of $Q$-component
spins. In particular, albeit two lattice sites belonging to the same
FK cluster will necessarily have the same spin value, the converse is
not true. It would therefore seem more natural to perform the
expansion in powers of ${\rm e}^K$ and consider as basic geometrical
observables the connected domains with a constant value of the Potts
spin, henceforth referred to as {\em spin clusters}.

The properties of these spin clusters have as a rule remained ill
understood. This is particularly frustrating, since those are the very
clusters that one would observe in an actual experiment on a magnetic
alloy in the Potts universality class. The case of the Ising model
$Q=2$ is an exception to this rule
\cite{DuplantierSaleur,Vanderzande}, but this is due to its
``coincidental'' equivalence to the O($n$) vector model with
$n=1$. Indeed, defining the Ising spins on the triangular lattice, the
corresponding O($n$) model is described by self and mutually avoiding
loops on the hexagonal lattice, and such loops are readily treated by
CFT and SLE techniques.

\begin{figure}
\begin{center}
 \includegraphics[width=0.3\textwidth]{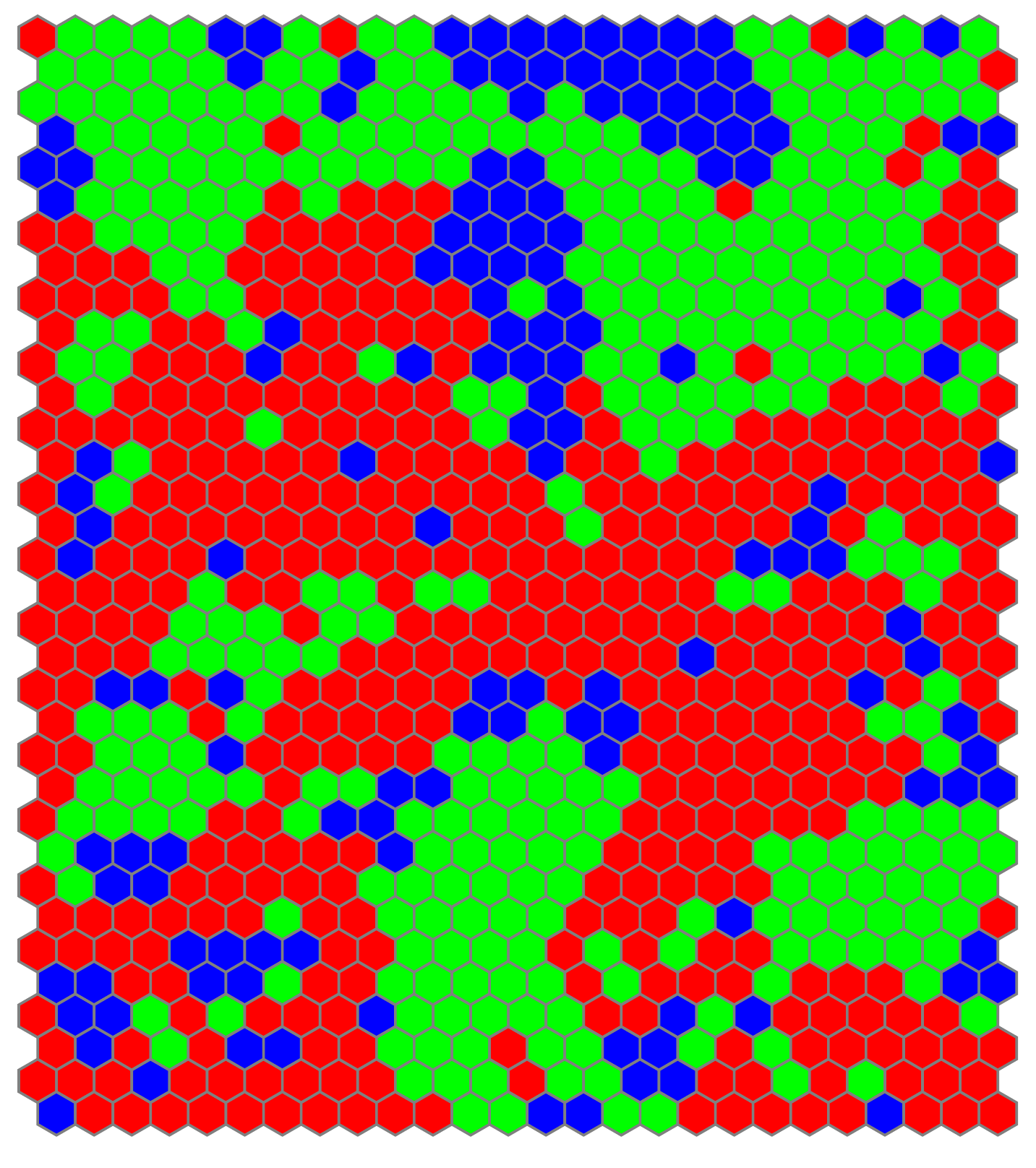}
 \includegraphics[width=0.3\textwidth]{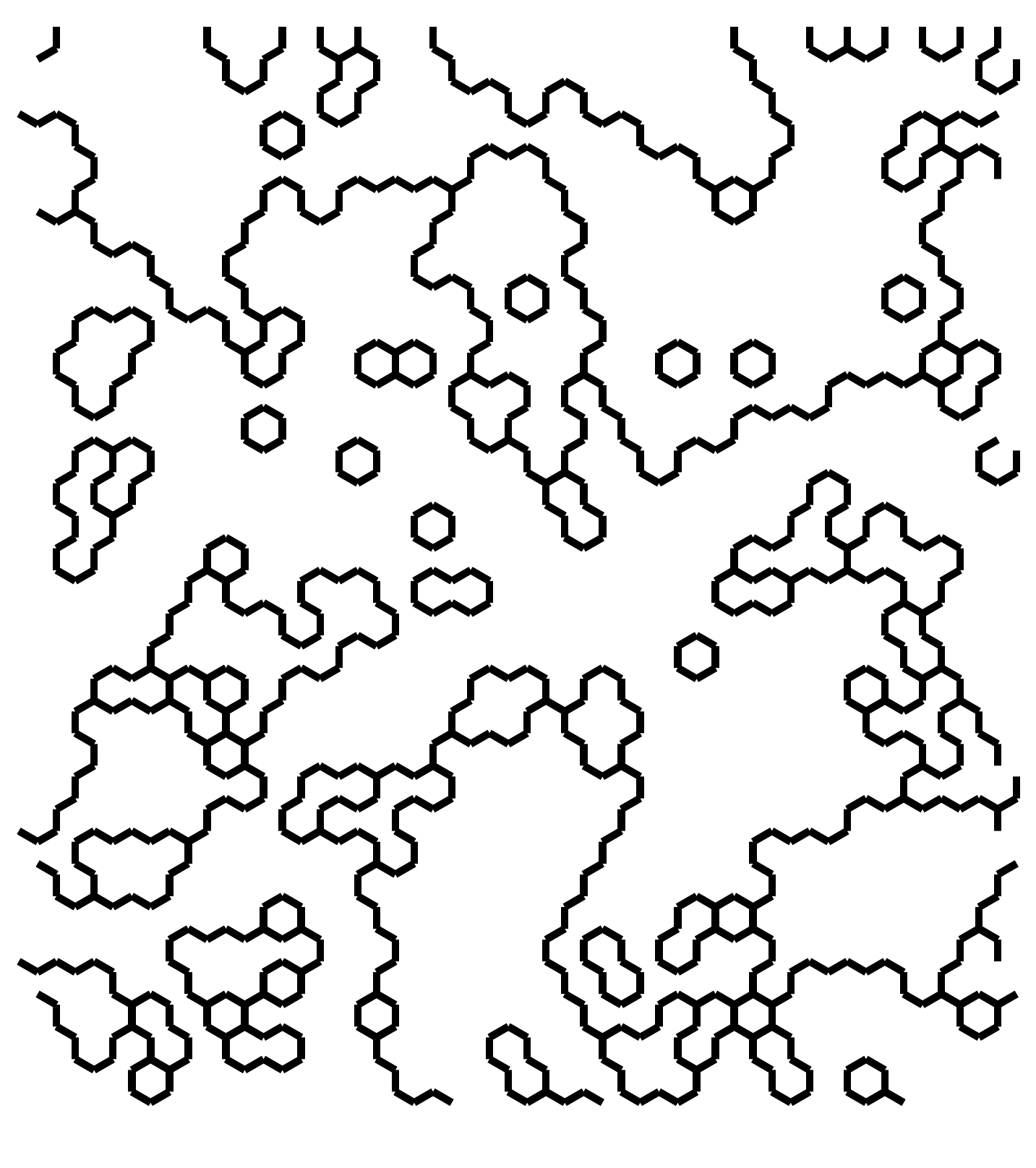}
\end{center}
 \caption{A configuration of the $Q=3$ Potts model, and the
   corresponding set of branching domain walls.}
 \label{fig:pottsDW}
\end{figure}

For general $Q$, the salient feature of Potts spin clusters is that
the domain walls separating different clusters undergo branchings and
crossings (see Fig.~\ref{fig:pottsDW}). These phenomena are however
absent for $Q=2$ with the above choice of lattice. It is precisely
these branchings and crossings that make the application of exact
techniques---such as CG mappings or Bethe ansatz
diagonalisation---very difficult, if not impossible. The belief that
spin clusters are indeed conformally invariant for other values of $Q
\neq 2$ in the critical regime has even been challenged at times, but
seems however well established by now \cite{Coniglio,Qian}.

Some progress has been accomplished in the $Q=3$ case
\cite{Vanderzande, Janke} by speculating that the spin clusters in the
critical Potts model would be equivalent to FK clusters in the {\em
  tricritical} Potts model \cite{Stella,Janke}. This
equivalence has however not been proven, and is moreover restricted so
far to the simplest geometrical questions
\cite{Duplantierduality}. The equivalence can also be understood as a
relationship with the dilute O($n$) model \cite{CardyGamsa}.

The Potts model has been used recently to build a new class of 2D
quantum lattice models that exhibit topological order
\cite{FendleyModels}. Both FK clusters and domain walls between spin
clusters are important in these models.  We should also mention that
spin domain walls have been studied numerically---and their difference
from the FK clusters pointed out---in the context of the $Z_N$
parafermionic models \cite{Picco_Santachiara}.

Apart from issues of branching and crossing, another major hurdle in
the study of Potts spin clusters has come from the lack of a
formulation that can be conveniently extended to $Q$ a real
variable. In the case of loops surrounding the FK clusters, this
formulation led naturally to the introduction of powerful algebraic
tools via the Temperley-Lieb (TL) algebra, and to the equivalence with
the six-vertex model---the eventual key to the exact solution of the
problem \cite{JJreview}. Factors of $Q$ then appear naturally through
a parameter in the TL algebra, or---via a geometrical
construction---as complex vertex weights in the six-vertex
formulation. Also for spin clusters can $Q$ be promoted to an
arbitrary variable: the weight of a set of spin clusters is simply the
chromatic polynomial of the graph dual to the domain walls. From the
point of view of the TL algebra, the domain walls are composite
(spin-1) objects, hence more complicated than the FK clusters and
loops.  Recent work on the related Birman-Wenzl-Murakami (BWM) algebra
\cite{ReadFendley,Fendley} suggests that this formulation might be
amenable to the standard algebraic and Bethe ansatz techniques,
although such a lofty goal has not been achieved so far.

In this paper we report major progress towards the understanding of
Potts spin clusters, both in the bulk and the boundary case. A brief
account on the bulk case has recently appeared elsewhere \cite{Letter}.
Our results are of two kinds. On the one hand, we
develop a transfer matrix technique which allows the formulation and
numerical study of the spin clusters for all real $Q$. On the other
hand, we identify the geometrical events that give rise to conformal
correlations, and provide exact (albeit numerically determined)
expressions for infinite families of critical exponents, similar to the
familiar ``$L$-legs'' exponents \cite{JJreview,Nienhuis} for TL
loops. Surprisingly, we find that geometrical properties of spin
clusters encompass all integer indices $(r,s)$ in the Kac table
$h_{r,s}$. An analytical derivation of our results appears for now
beyond reach, in part because the algebraic properties of our transfer
matrix are still ill understood. We do however provide some exact
results based on an approach which does not involve a CG mapping, but
rather the use of a massless scattering description.

Apart from the bulk critical exponents \cite{Letter}, we report here
the boundary critical exponents for free, fixed, and mixed boundary
conditions. This means that the Potts spins on a segment of the
boundary are restricted to take $Q_1$ values, where $Q_1$ can assume
any real value and is in general different from the number of states
$Q$ taken by bulk spins. These results can be viewed as a further step
in the programme \cite{CBL} of classifying non-unitary boundary
conditions in 2D geometrical models.

\section{Domain wall expansion}

\begin{figure}
 \begin{center}
 \includegraphics[width=0.3\textwidth]{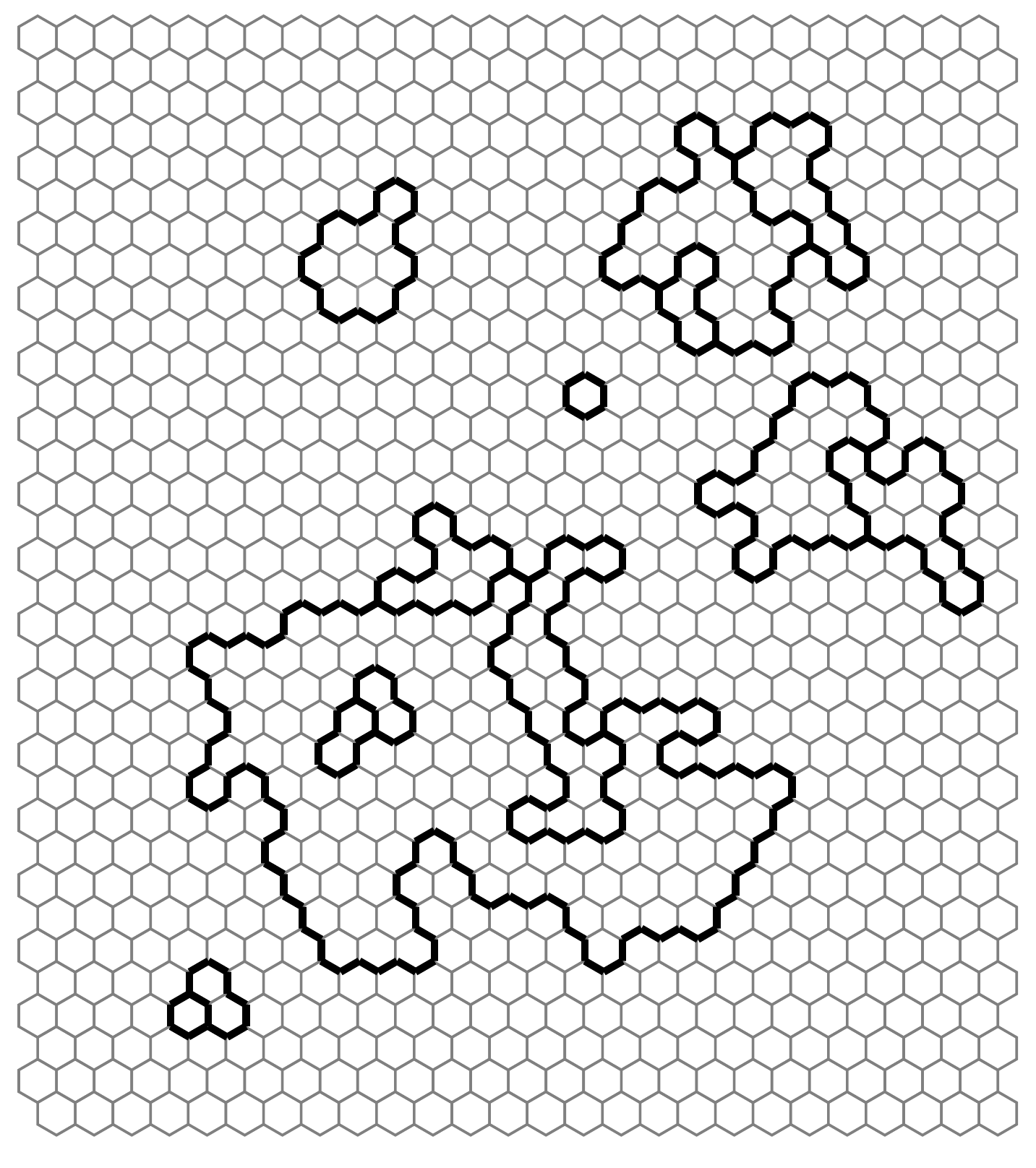}
 \includegraphics[width=0.3\textwidth]{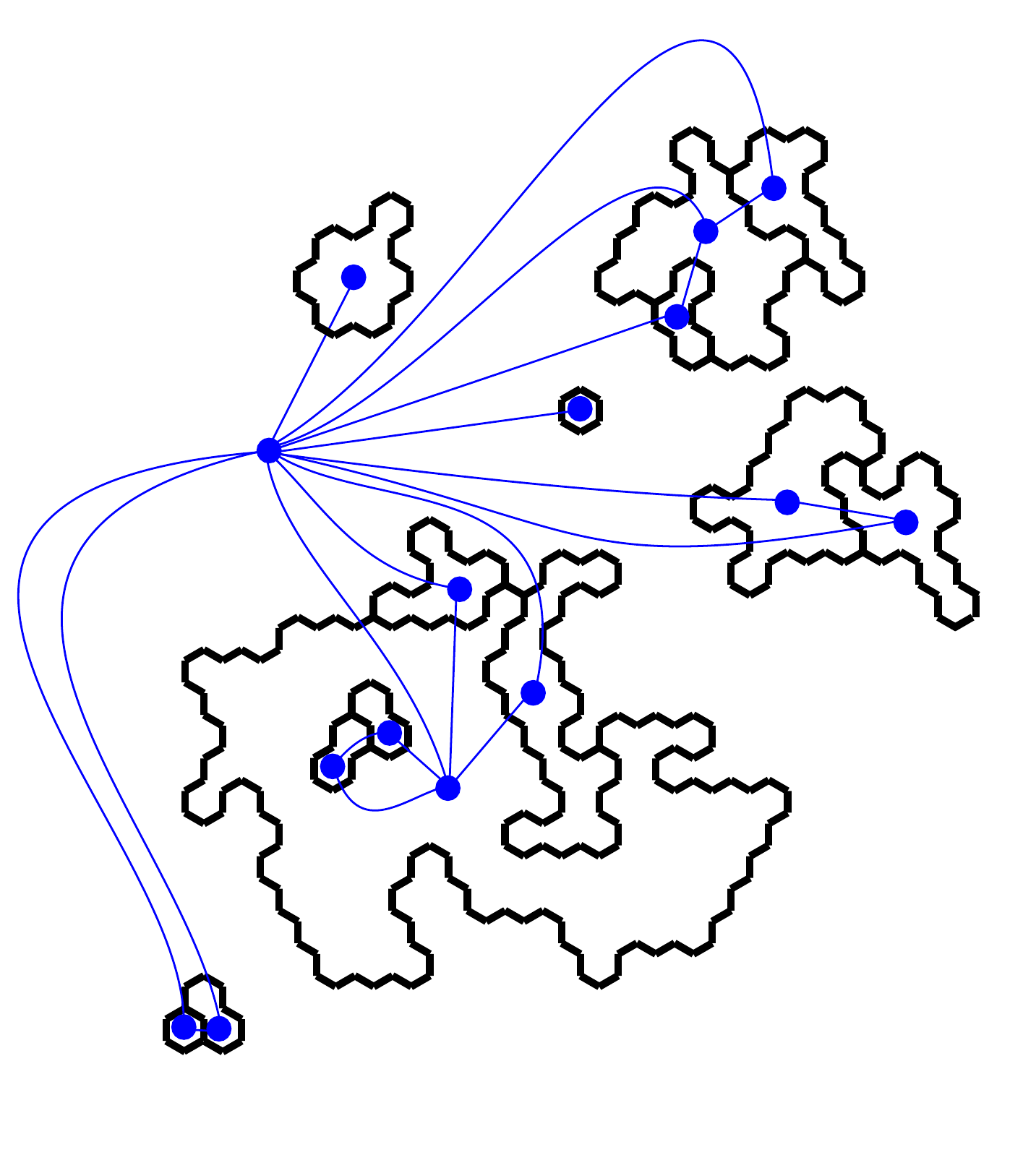}
 \end{center}
 \caption{A domain-wall configuration corresponding to a graph $G$,
   and its dual graph $\hat{G}$.}
 \label{fig:DWgraph}
\end{figure}

The domain wall (DW) expansion of (\ref{Potts}) involves all possible
configurations of domain walls that can be drawn on the dual of ${\cal
  L}$ (see Fig.~\ref{fig:DWgraph}). A DW configuration is given by a
graph $G$ (not necessarily connected). The faces of $G$ are the spin
clusters.  Since we do not specify the colour of each of these
clusters, a DW configuration has to be weighted by the chromatic
polynomial $\chi_{\hat{G}}(Q)$ of the dual graph $\hat{G}$.  Initially
$\chi_{\hat{G}}(Q)$ is defined as the number of colourings of the
vertices of the graph $\hat{G}$, using colours $\{1,2,\ldots,Q\}$, with
the constraint that neighbouring vertices have different colours.  This
is indeed a polynomial in $Q$ for any $G$, and so can be evaluated for
any real $Q$ (but $\chi_{\hat{G}}(Q)$ is integer only when $Q$ is
integer).  For example, the chromatic polynomial of the graph
$\hat{G}$ on Fig.~\ref{fig:DWgraph} is $Q (Q-1)^7 (Q-2)^7$.
The partition function (\ref{Potts}) can thus be written as a sum over all
possible DW configurations 
\be
 Z = {\rm e}^{N K} \sum_{G} \, \left( {\rm e}^{-K} \right)^{{\rm length}(G)}
  \, \chi_{\hat{G}}(Q)
 \label{PottsDW}
\ee
where $N$ is the number of spins, and ${\rm length}(G)$ denotes the
total length of the domain walls.

\begin{figure}[htbp]
 \begin{center}
 a. \,\includegraphics[width=0.4\textwidth]{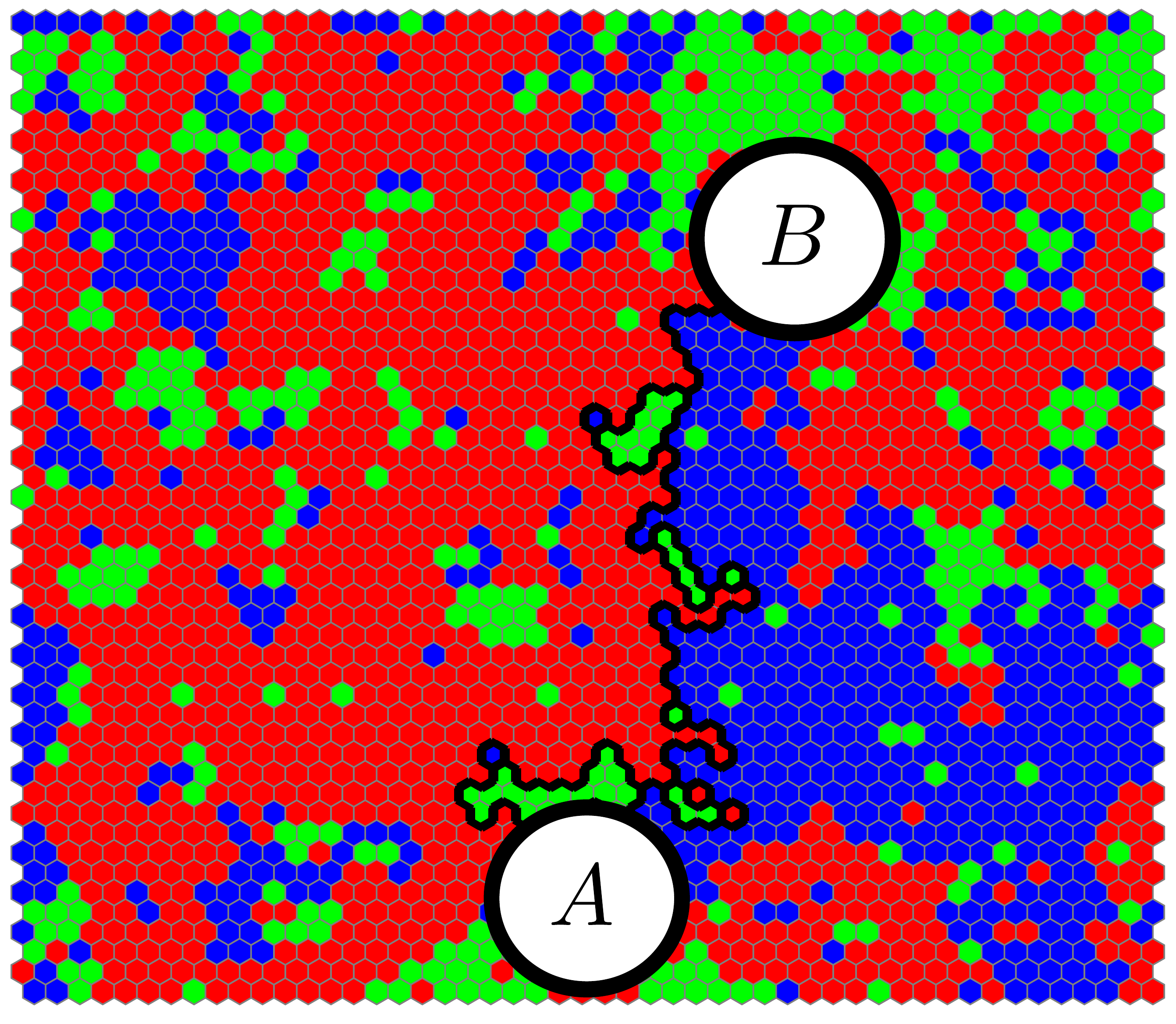}
    \includegraphics[width=0.4\textwidth]{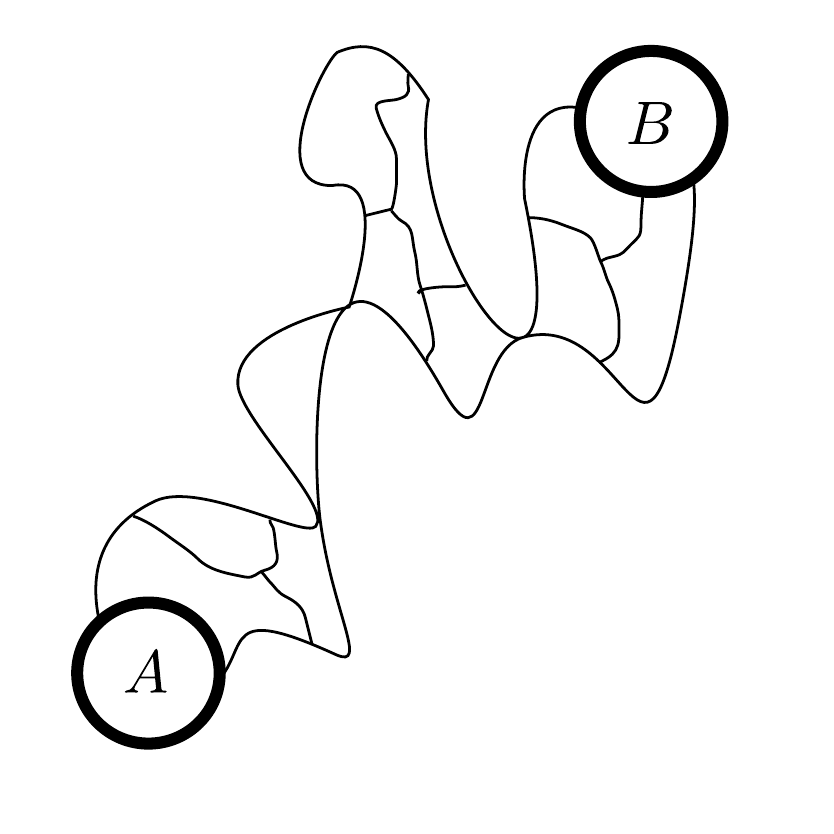} \\ \vspace{0.3cm}
 b. \, \includegraphics[width=0.4\textwidth]{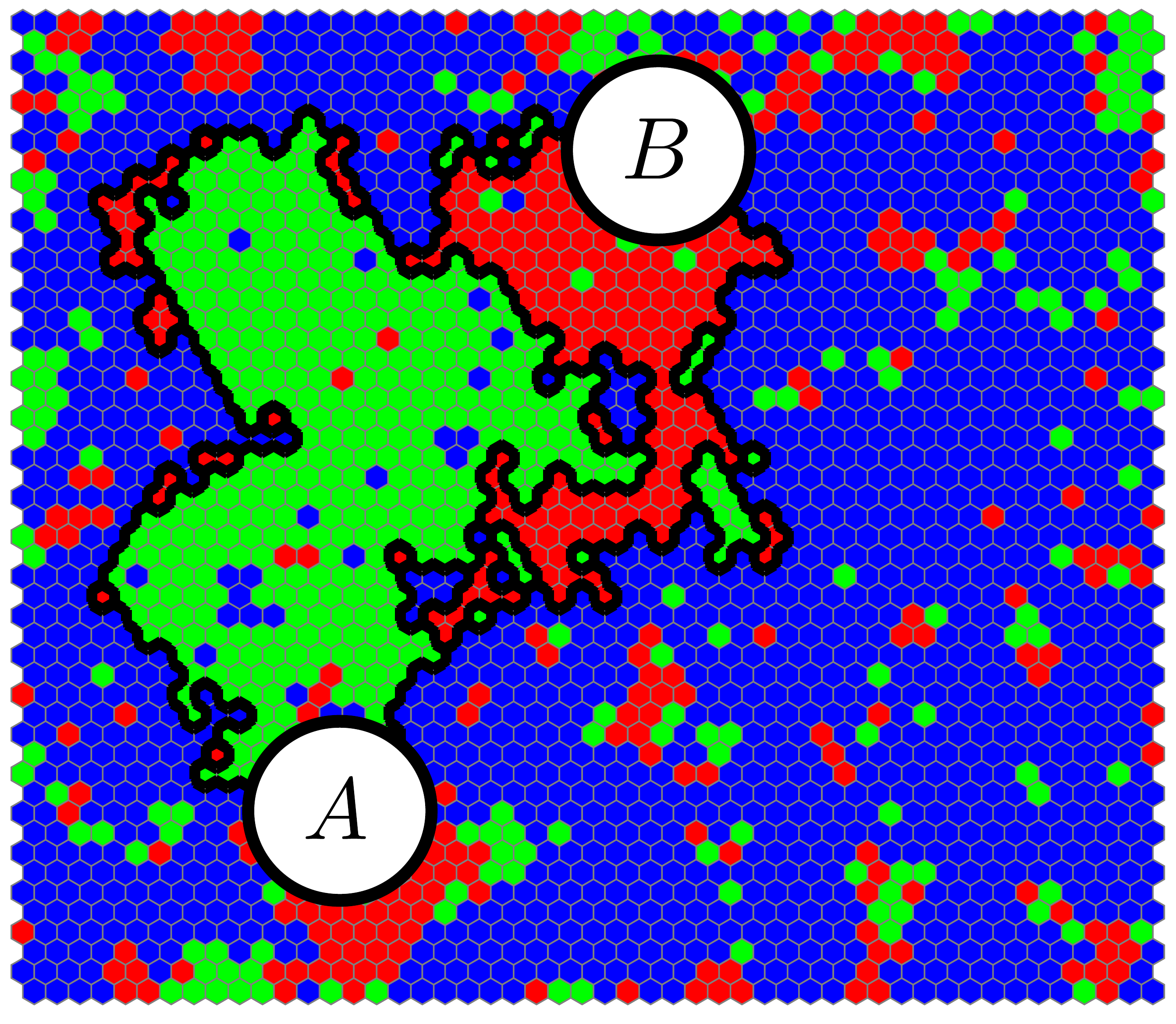}
    \includegraphics[width=0.4\textwidth]{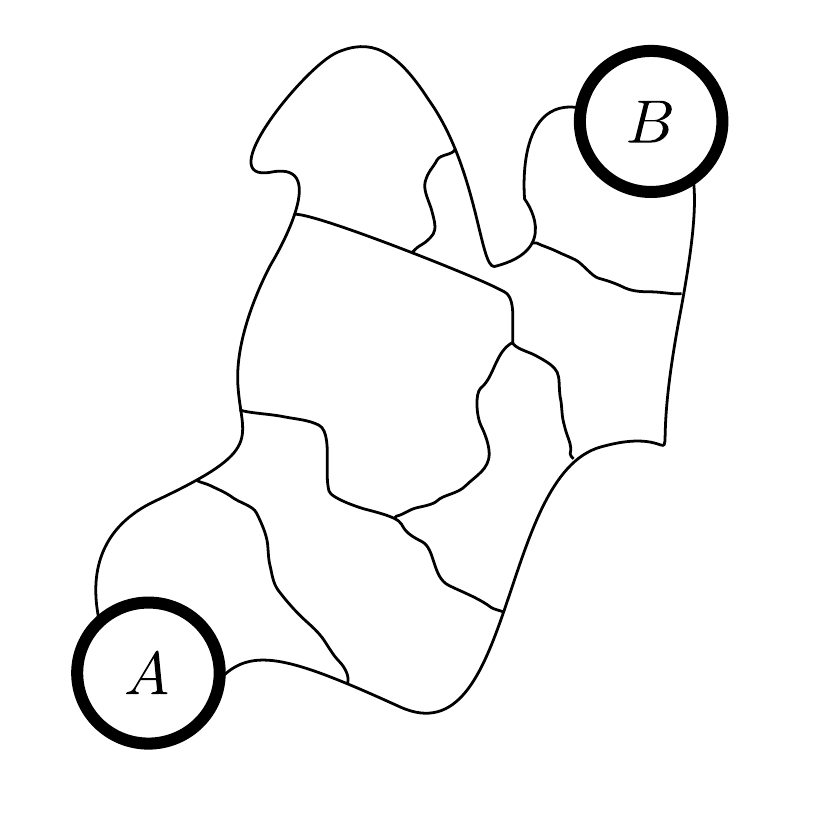}
 \end{center}
 \caption{The two different types of DW, here shown in the bulk case
   (geometry of the infinite plane). A \textit{thin} DW
   corresponds to the interface between two clusters of different
   colours (a), while for a \textit{thick} DW the two clusters have the
   same colour. An illustration for the $Q=3$ Potts model is given
   (left) as well as a schematic picture for non-integer $Q$ (right).}
 \label{DW2types}
\end{figure}

The fundamental geometric object we consider is a connected part of a
domain wall that separates two clusters. One can ask how the
probability, that a certain number of such DW connect a small (in
units of the lattice spacing) neighbourhood $A$ to another small
neighbourhood $B$, decays when the distance $x$ between $A$ and $B$
increases. Each DW separates two spin clusters which connect $A$ and
$B$.  There are in fact two types of such DW, depending on the
relative colouring of the two clusters that are separated. If the two
clusters have different colours, they can touch, so the DW is
\textit{thin} (see Fig.~\ref{DW2types}.a).  If the two clusters have
the same colour, then they cannot touch (otherwise they would not be
distinct), so the DW has to be \textit{thick} (see
Fig.~\ref{DW2types}.b).

\begin{figure}[htbp]
 \begin{center}
  \includegraphics[width=0.49\textwidth]{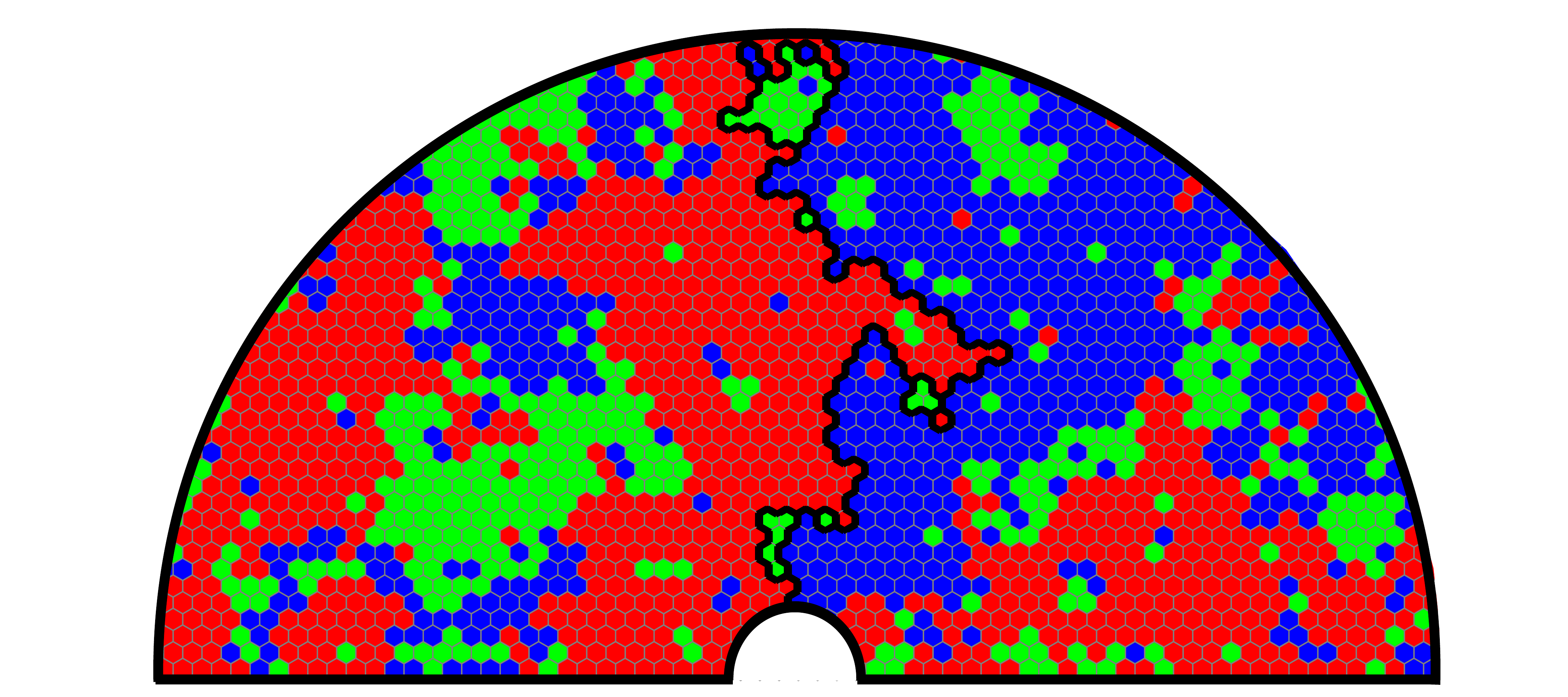}
  \includegraphics[width=0.49\textwidth]{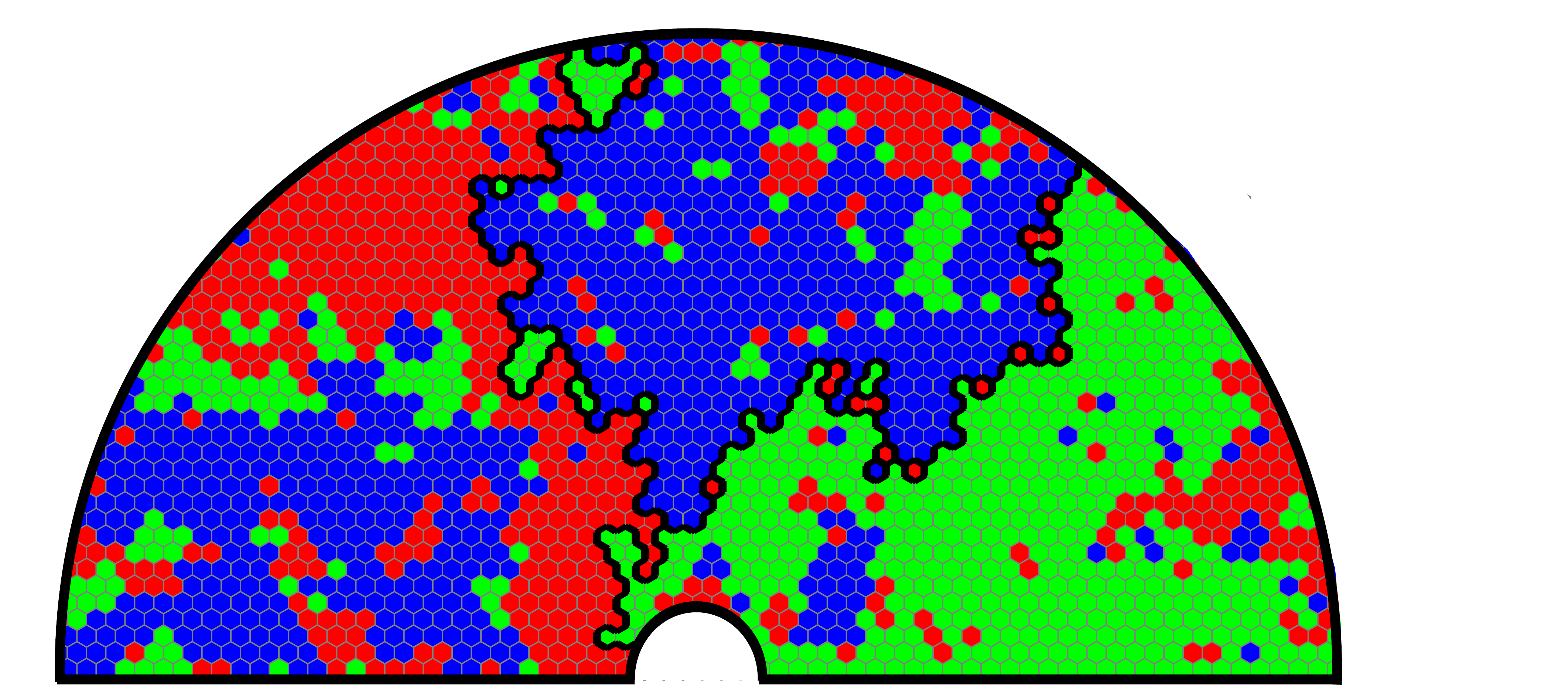}
 \end{center}
 \caption{A thin DW in the half plane geometry for $Q=3$. The left
   (resp.\ right) panel illustrates the case where the positive real
   axis supports free (resp.\ fixed) boundary conditions.}
 \label{halfplane12}
\end{figure}

Several different geometries are of interest. In the bulk case, the
neighbourhoods $A$ and $B$ are at arbitrary, but widely separated,
locations in the infinite plane. This is conformally equivalent to an
infinitely long cylinder---a strip with periodic boundary
conditions---with $A$ and $B$ situated at the two extremities.  In the
boundary case, the geometry is that of the upper half plane, with $A$
being at the origin and $B$ far away from the real axis.  The boundary
conditions are taken to be free, fixed or mixed on the positive real
axis, and free on the negative real axis. In other words, the Potts
spins can take $Q_1$ (resp.\ $Q$) values on the positive
(resp.\ negative) real axis.  This geometry is illustrated in
Fig.~\ref{halfplane12}; it is conformally equivalent
to an infinitely long strip, with $A$ and $B$ situated at respectively
the upper and lower extremities, and spins on the left (resp.\ right)
rim taking $Q_1$ (resp.\ $Q$) values. We shall use the cylinder and
strip geometries below to conduct our numerical calculations.

We can now state the central claims of this paper. Consider the 2D
Potts model for any real $Q$ in the critical regime $0 \le Q \le
4$. Then the probability $P$ that the two regions $A$ and $B$, with
separation $x \gg 1$, are connected by $\ell_1$ thin DW and $\ell_2$
thick DW decays algebraically. In the bulk case (geometry of the
plane) the corresponding critical exponent is denoted $h(Q, \ell_1,
\ell_2)$, and we have $P \propto x^{-4 h(Q,\ell_1
  ,\ell_2)}$. Equivalently, on a long cylinder of size $L \times \ell$
with $\ell \gg L$, and $A$ and $B$ identified with the opposite ends
of the cylinder, the decay is exponential: $P \propto {\rm e}^{- 4 \pi
  (\ell/L) \, h(Q, \ell_1, \ell_2)}$. In the boundary case (geometry
of the half plane) the critical exponent for free boundary conditions
is denoted $\widetilde{h}(Q,\ell_1,\ell_2)$.

Below we check these assertions numerically, and we observe that the
numerical values of the exponents match the formulae
\begin{eqnarray}
 h(Q, \ell_1, \ell_2) &=& h_{\ell_1 - \ell_2, 2 \ell_1} \,, \nonumber \\
 \widetilde{h}(Q,\ell_1,\ell_2) &=& h_{1+2(\ell_1-\ell_2),1+4 \ell_1} \,,
 \label{mainconj}
\end{eqnarray}
where we have used the Kac parametrisation of CFT
\begin{equation}
 h_{r,s} = \frac{\left( r-s \, \kappa/4 \right)^2 - \left( 1 - \kappa/4 \right)^2}{\kappa}
 \label{Kac}
\end{equation}
and $2 \le \kappa \le 4$ parametrises $Q = 4\left( \cos \frac{\kappa
    \,\pi}{4} \right)^2 \in [0,4]$.


It remains to give our results for the mixed boundary conditions with
parameter $Q_1$ (of which fixed boundary conditions corresponds to the
special case $Q_1 = 1$). These follow by conformal fusion of the
free-to-mixed boundary condition changing operator $\Phi_{r_0,s_0}$
that has been worked out in \cite{CBL} with the operator
$\Phi_{1+2(\ell_1-\ell_2),1+4 \ell_1}$ that inserts the required
number of DW with free boundary conditions. Details about this will
be given below in section~\ref{sec:mixed}.

We close this section with two remarks about the precise interpretation
of the main result (\ref{mainconj}).

First, in the expression (\ref{mainconj}) for the bulk exponent $h(Q,
\ell_1, \ell_2)$, the case $(\ell_1,\ell_2)=(0,1)$ of a single DW
(which must then necessarily be thick, due to the periodic boundary
conditions) is special. In that case (\ref{mainconj}) remains valid
only if the DW is forbidden to wrap around the neighbourhoods $A$ and
$B$ (in the geometry of the plane), or around the periodic direction
(in the equivalent cylinder geometry). Without that restriction being
imposed, we obtain another result:
\begin{equation}
 h(Q,0,1) = h_{0,1/2} \qquad
 \mbox{(wrapping allowed)} \,.
 \label{magn_exp}
\end{equation}

Second, in the boundary case the numbers $\ell_1$ and $\ell_2$
appearing in (\ref{mainconj}) must be defined more carefully. Indeed,
when there are $\ell \equiv \ell_1+\ell_2$ propagating spin clusters,
only the nature (`thin' or `thick') or the $\ell-1$ DW separating them
is clear by the above definition, whereas it is not yet clear how to
characterise the `half domain walls' that separate the two outermost
spin clusters from the boundaries. The correct and unambiguous
definition of $\ell_1$ and $\ell_2$ is as follows: Let the leftmost
spin cluster contribute one unit to $\ell_1$, and let each of the
$\ell-1$ subsequent spin clusters contribute one unit to $\ell_1$
(resp.\ $\ell_2$) if it has a different (resp.\ the same) colour as
the cluster on its left.

\section{Transfer matrix formulation}

The DW expansion (\ref{PottsDW}) may appear unwieldy and difficult to
study numerically for non-integer $Q$.  There nevertheless exists
several Monte Carlo methods for studying the Potts model when $Q$ is
not an integer \cite{Sweeny,Chayes_Machta,Deng_Blote}, all of which
are roughly speaking based on the FK cluster representation. It is
possible to reconstruct the spin clusters from the Chayes-Machta
algorithm \cite{Chayes_Machta} by performing the bond-adding step at
zero temperature. Following our brief account \cite{Letter}, this
method has very recently been used \cite{Zatelepin_Shchur} to verify a
special case of (\ref{mainconj}) corresponding to $h(Q,0,1) =
h_{-1,0}$ (with wrapping forbidden).

As in \cite{Letter} we shall however take a quite different route and
resort to a transfer matrix construction. This has the advantage of
linking up more easily with CFT \cite{CardyBloteNightingale}, and
giving very precise numerical results for all the boundary conditions
outlined above. Moreover, the ensuing transfer matrix formalism is no
more complicated than the one \cite{Blote82} routinely used in the
study of the FK clusters.

\subsection{State space}

Consider a strip of the square lattice of width $L$ spins (boundary
conditions will be detailed later). The basis states on which the
row-to-row transfer matrix $T$ acts contain one colour label $c_i$ per
spin.  By definition, one has $c_i = c_j$ if and only if $\sigma_i =
\sigma_j$ (i.e., the two spins on sites $i$ and $j$ have the same
colour). The colour labels $c_i$ contain less information than the
spin colours $\sigma_i$ themselves. For instance, any configuration in
which the first and third spins have the same colour, no matter which
one, and no other spins have identical colours, is represented by
\begin{equation}
\label{eq:state}
\begin{tikzpicture}
        \filldraw (0,0) circle (0.7mm) node[above]{$c_1$};
        \filldraw (0.5,0) circle (0.7mm) node[above]{$c_2$};
        \filldraw (1,0) circle (0.7mm) node[above]{$c_1$};
        \filldraw (1.5,0) circle (0.7mm) node[above]{$c_4$};
        \draw (2,0) node{$\dots$};
        \filldraw (2.5,0) circle (0.7mm) node[above]{$c_L$};
\end{tikzpicture}
\end{equation}
With these conventions we are always able to recognise whether to
spins have the same colour or not, even if we do not know the precise
value of this colour. But this is all that is needed to determine
the Boltzmann weights in (\ref{Potts}).

For a row of $L=3$ vertices, and for any (non-integer) $Q$, there
are precisely five basis states
\begin{equation}
\label{basisL3}
 \begin{tikzpicture}
        \filldraw (0,0) circle (0.7mm) node[above]{$c_1$};
        \filldraw (0.4,0) circle (0.7mm) node[above]{$c_1$};
        \filldraw (0.8,0) circle (0.7mm) node[above]{$c_1$};
\end{tikzpicture}, \, \begin{tikzpicture}
        \filldraw (0,0) circle (0.7mm) node[above]{$c_1$};
        \filldraw (0.4,0) circle (0.7mm) node[above]{$c_2$};
        \filldraw (0.8,0) circle (0.7mm) node[above]{$c_1$};
\end{tikzpicture}, \, \begin{tikzpicture}
        \filldraw (0,0) circle (0.7mm) node[above]{$c_1$};
        \filldraw (0.4,0) circle (0.7mm) node[above]{$c_1$};
        \filldraw (0.8,0) circle (0.7mm) node[above]{$c_2$};
\end{tikzpicture} , \, \begin{tikzpicture}
        \filldraw (0,0) circle (0.7mm) node[above]{$c_1$};
        \filldraw (0.4,0) circle (0.7mm) node[above]{$c_2$};
        \filldraw (0.8,0) circle (0.7mm) node[above]{$c_2$};
\end{tikzpicture}, \, \begin{tikzpicture}
        \filldraw (0,0) circle (0.7mm) node[above]{$c_1$};
        \filldraw (0.4,0) circle (0.7mm) node[above]{$c_2$};
        \filldraw (0.8,0) circle (0.7mm) node[above]{$c_3$};
\end{tikzpicture}
\end{equation} 
Note that the last state would carry zero weight for $Q=2$, but
apart from that the number of basis states for any given $L$ will
be finite and independent of $Q$.

In general, the number of states is equal to the Bell number $B_L$ of
partitions of $L$ objects, with exponential generating function
\begin{equation}
 \sum_{L=0}^\infty \frac{B_L}{L!} x^L = \exp({\rm e}^x-1) \,.
\end{equation}
Note that $B_L$ differs from the Catalan number $C_L$ of {\em planar}
partitions, familiar from the FK cluster representation. In
particular, $\{1,2,1,2\}$ is a valid state for $L=4$.  For $L \gg 1$
one has asymptotically $C_L \sim 4^L$, whereas $B_L$ grows
super-exponentially. When $Q$ is integer, the number of states
truncates to $\approx Q^L / Q!$.

Let us take the time evolution direction to be upwards, so that
horizontal (resp.\ vertical) edges are space-like (resp.\ time-like).
We can write $T$ as a product of elementary transfer matrices, each
represented symbolically as a rhombus surrounding a single lattice
edge, and corresponds to the addition of that edge to the lattice.
This edge links spins (shown as solid circles) on diametrically
opposite sites of the rhombus. On an $L=4$ square lattice with
periodic boundary conditions this reads
$$T = \begin{tikzpicture}[scale=0.5]
        \draw (0,1) -- (0,0) -- (1,1) -- (2,0) -- (3,1) -- (4,0) -- (4,1) -- (3,0) -- (2,1) -- (1,0) -- (0,1) -- (0.5,1.5) -- (1,1) -- (1.5,1.5) -- (2,1) -- (2.5,1.5) -- (3,1) -- (3.5,1.5) -- (4,1);
        \filldraw (0,0) circle (1.4mm);
        \filldraw (1,0) circle (1.4mm);
        \filldraw (2,0) circle (1.4mm);
        \filldraw (3,0) circle (1.4mm);
        \filldraw (4,0) circle (1.4mm);
        \filldraw (0,1) circle (1.4mm);
        \filldraw (1,1) circle (1.4mm);
        \filldraw (2,1) circle (1.4mm);
        \filldraw (3,1) circle (1.4mm);
        \filldraw (4,1) circle (1.4mm);
\end{tikzpicture} \,.$$

A rhombus
$\begin{tikzpicture}[scale=0.18] \draw (0,0) -- (1,1) -- (2,0) -- (1,-1) -- cycle;
        \filldraw (1,1) circle (2.4mm);
        \filldraw (1,-1) circle (2.4mm);
\end{tikzpicture}$
corresponds to a vertical edge, and acts on a basis state $s$ as
follows. If exactly $Q_s$ distinct colour labels $\{c_k\}$ are used in
$s$, then the new colour label $c_i'$ of the spin $\sigma_i$ can be
either unchanged $c_i'=c_i$ (with weight ${\rm e}^K$), or any one of the
other labels already in use $c_i'=c_k$ (each with weight $1$), or a
new one $c_i' \notin \{c_k\}$ (with weight $Q-Q_s$). Note that this
latter weight is in general non-integer, and is responsible for the
correct computation of the chromatic polynomial $\chi_{\hat{G}}(Q)$.
On an example with $Q_s=3$ this reads explicitly
$$
\begin{tikzpicture}
	\filldraw (0,0) circle (0.7mm) node[below]{$c_1$};
	\filldraw (0.5,0) circle (0.7mm) node[below]{$c_2$};
	\filldraw (1,0) circle (0.7mm) node[below]{$c_3$};
	\filldraw (1,0.5) circle (0.7mm) node[above]{$c_3'$};
	\draw (1,0) -- (1.25,0.25) -- (1,0.5) -- (0.75,0.25) -- cycle;
	\draw (1.75,0.25) node{$=$}; 
	\begin{scope}[xshift=3.5cm,yshift=0.3cm]
		\draw (-0.8,0) node{$\delta_{c_3,c_3'}\, {\rm e}^K$};
		\filldraw (0,0) circle (0.7mm) node[below]{$c_1$};
		\filldraw (0.4,0) circle (0.7mm) node[below]{$c_2$};
		\filldraw (0.8,0) circle (0.7mm) node[below]{$c_3$};
	\end{scope}
	\begin{scope}[xshift=6.2cm,yshift=0.3cm]
		\draw (-0.8,0) node{$+\; \delta_{c_1,c_3'}$};
		\filldraw (0,0) circle (0.7mm) node[below]{$c_1$};
		\filldraw (0.4,0) circle (0.7mm) node[below]{$c_2$};
		\filldraw (0.8,0) circle (0.7mm) node[below]{$c_1$};
	\end{scope}
	\begin{scope}[xshift=8.7cm,yshift=0.3cm]
		\draw (-0.8,0) node{$+ \;\delta_{c_2,c_3'}$};
		\filldraw (0,0) circle (0.7mm) node[below]{$c_1$};
		\filldraw (0.4,0) circle (0.7mm) node[below]{$c_2$};
		\filldraw (0.8,0) circle (0.7mm) node[below]{$c_2$};
	\end{scope}
	\begin{scope}[xshift=8.7cm,yshift=-1.0cm]
		\draw (-3.7,0) node{$+\; \left( 1-\delta_{c_1,c_3'}- \delta_{c_2,c_3'} - \delta_{c_3,c_3'} \right) \left( Q-Q_s\right)$};
		\filldraw (0,0) circle (0.7mm) node[below]{$c_1$};
		\filldraw (0.4,0) circle (0.7mm) node[below]{$c_2$};
		\filldraw (0.8,0) circle (0.7mm) node[below]{$c_3$};
	\end{scope}
\end{tikzpicture}
$$
The last state could be written
$\begin{tikzpicture} \filldraw (0,0) circle (0.7mm) node[above]{$c_1$};
		\filldraw (0.4,0) circle (0.7mm) node[above]{$c_2$};
		\filldraw (0.8,0) circle (0.7mm) node[above]{$c_3'$};
\end{tikzpicture}$,
but this is equivalent to
$\begin{tikzpicture} \filldraw (0,0) circle (0.7mm) node[above]{$c_1$};
		\filldraw (0.4,0) circle (0.7mm) node[above]{$c_2$};
		\filldraw (0.8,0) circle (0.7mm) node[above]{$c_3$};
\end{tikzpicture}$
in the basis (\ref{basisL3}).

A rhombus
$\begin{tikzpicture}[scale=0.18] \draw (0,0) -- (1,1) -- (2,0) -- (1,-1)
 -- cycle;
	\filldraw (0,0) circle (2.4mm);
	\filldraw (2,0) circle (2.4mm);
\end{tikzpicture}$
adding a horizontal edge between vertices $i$ and $i+1$ corresponds
simply to a diagonal matrix, with a weight ${\rm e}^K$ if $c_i=c_{i+1}$, and
$1$ otherwise.

With these rules at hand, one can write the periodic (cylinder
geometry) $L=3$ transfer matrix for arbitrary $Q$ in the basis
(\ref{basisL3}) as an instructive example:
$T \, =\, h_1 \cdot h_2 \cdot h_3 \cdot v_1 \cdot v_2 \cdot v_3$ with
\begin{equation*}
	v_1  =  \left(\begin{array}{ccccc} 
		{\rm e}^K & 0 & 0 & 1 & 0 \\
		0 & {\rm e}^K & 1 & 0 & 1 \\
		0 & 1 & {\rm e}^K & 0 & 1 \\
		Q-1 & 0 & 0 & {\rm e}^K +Q-2 & 0 \\
		0 & Q-2 & Q-2 & 0 & {\rm e}^K + Q-3 \\
 \end{array}  \right) 
\end{equation*}
and $h_1 = {\rm diag}({\rm e}^K,1, {\rm e}^K,1,1)$.
The remaining matrices can be obtained from those given by cyclic
permutations of the colour labels:
\begin{equation*}
	v_2  =  \left(\begin{array}{ccccc} 
		{\rm e}^K & 1 & 0 & 0 & 0 \\
		Q-1 & {\rm e}^K+Q-2 & 0 & 0 & 0 \\
		0 & 0 & {\rm e}^K & 1 & 1 \\
		0 & 0 & 1 & {\rm e}^K & 1 \\
		0 & 0 & Q-2 & Q-2 & {\rm e}^K + Q-3 \\
 \end{array}  \right) 
\end{equation*}
\begin{equation}
	\label{TL3}
	v_3  =  \left(\begin{array}{ccccc} 
		{\rm e}^K & 0 & 1 & 0 & 0 \\
		0 & {\rm e}^K & 0 & 1 & 1 \\
		Q-1 & 0 & {\rm e}^K+Q-2 & 0 & 0 \\
		0 & 1 & 0 & {\rm e}^K & 1 \\
		0 & Q-2 & 0 & Q-2 & {\rm e}^K + Q-3 \\
 \end{array}  \right) 
\end{equation}
and $h_2 = {\rm diag}({\rm e}^K,1,1,{\rm e}^K,1)$, $h_3 = {\rm
  diag}({\rm e}^K,{\rm e}^K,1,1,1)$.

To get a non-periodic strip geometry instead of a periodic cylinder,
one needs simply to omit the last horizontal edge which is responsible
for the periodic boundary conditions. For instance, with $L=3$ one
would have $T \, =\, h_1 \cdot h_2 \cdot v_1 \cdot v_2 \cdot v_3$.
This corresponds to free boundary conditions.  Mixed boundary
conditions are obtained by constraining the spins on the left boundary
of the strip to belong to a subset of $Q_1$ different colours, with $0
< Q_1 \le Q$. Spins on the right boundary remain free. For $Q_1$
integer this can be coded in the transfer matrix by considering that
the number of colours in use in a given state is always at least $Q_1$
(so that in particular $Q_s \ge Q_1$).  These first $Q_1$ colours are
considered fixed, whereas other colours $Q_1+1,Q_1+2,\ldots$ are
defined only relative to the fixed ones, as described above. In
particular, the choice $Q_1 = 1$ corresponds to fixed boundary
conditions.

The leading eigenvalue $\Lambda_0$ of the transfer matrix $T$ with
periodic boundary conditions gives the ground state free energy $f_0 =
-\frac{1}{L} \log \Lambda_0$. This $f_0$ coincides precisely with that
of the usual FK transfer matrix \cite{Blote82}, even when $Q$ is
non-integer.%
\footnote{The reader may wish to check this fact on the $L=3$ example
  given above.}
Its finite-size corrections possess a universal $L^{-2}$ term whose
coefficient determines the central charge of the corresponding CFT
\cite{CardyBloteNightingale}.

\subsection{Correlation functions}

However, to obtain the desired two-point correlation functions of DW,
we need to construct a variant transfer matrix $T_{\ell_1,\ell_2}$
which imposes the propagation of $\ell_1$ thin and $\ell_2$ thick DW
along the time direction of the cylinder (or strip). From its leading
eigenvalue $\Lambda_{\ell_1,\ell_2}$ one can determine the energy gap
$\Delta f_{\ell_1,\ell_2} = -\frac{1}{L}
\log(\Lambda_{\ell_1,\ell_2} / \Lambda_0)$ whose finite-size scaling in
turn determines the critical exponents $h(Q, \ell_1, \ell_2)$ and
$\widetilde{h}(Q,\ell_1,\ell_2)$ \cite{CardyBloteNightingale,JJreview}.

To this end, the basis states (\ref{basisL3}) need to be endowed with
some additional information about the connectivity of the spin
clusters, ensuring the propagation of the desired number and types of
DW. The crucial point is that we need to know whether two spins having
the same colour label $c_i$ also belong to the same cluster. Thus the states
we use in the final transfer matrix have the form
\begin{equation}
	\begin{tikzpicture}
	\draw[line width=2pt,gray] (1,0) arc (-180:0:0.25 and 0.5); 
	\filldraw (0,0) circle (0.7mm) node[above]{$c_1$};
	\filldraw (0.5,0) circle (0.7mm) node[above]{$c_2$};
	\filldraw (1,0) circle (0.7mm) node[above]{$c_1$};
	\filldraw (1.5,0) circle (0.7mm) node[above]{$c_1$};
	\filldraw (2,0) circle (0.7mm) node[above]{$c_5$};
\end{tikzpicture}  \qquad \quad
	\begin{tikzpicture}
	\draw[line width=2pt,gray] (0,0) arc (-180:0:0.75 and 0.5); 
	\draw[line width=2pt,gray] (1,0) -- (0.75,-0.5); 
	\filldraw (0,0) circle (0.7mm) node[above]{$c_1$};
	\filldraw (0.5,0) circle (0.7mm) node[above]{$c_2$};
	\filldraw (1,0) circle (0.7mm) node[above]{$c_1$};
	\filldraw (1.5,0) circle (0.7mm) node[above]{$c_1$};
	\filldraw (2,0) circle (0.7mm) node[above]{$c_5$};
\end{tikzpicture} 
\label{linkings}
\end{equation}
meaning that two spins belong to the same spin cluster if and only if
they are linked up in this pictorial representation. Of course, only
spins with a common colour label can be linked up.  Thus, in the left
state of (\ref{linkings}), the spins on vertices $3$ and $4$ are in
the same cluster, but not in the same cluster as the spin $1$. In the
right state the spins $1$, $3$ and $4$ are all in the same
cluster. These two states are different. In the transfer matrix
evolution, each time two neighbour vertices correspond to the same
colour, the corresponding clusters are linked up.

Note that the possible ways of linking up the spins must respect
planarity.  For instance, in the $L=4$ state with colour labels
$\{1,2,1,2\}$, one cannot simultaneously link the first and third
spins, and the second and fourth, since this possibility is disallowed
by planarity.

To construct $T_{\ell_1,\ell_2}$ we modify the linked basis states
(\ref{linkings}) so that precisely $\ell_1 + \ell_2$ spin clusters are
{\em marked}. The colour labels of the marked clusters must respect
the chosen values of $\ell_1$ and $\ell_2$. In order to conserve the
marked clusters in the transfer matrix evolution, none of the marked
clusters must be ``left behind'', and two distinct marked clusters
must not be allowed to link up. When a marked and an unmarked cluster
link up, the result is a marked cluster.

To summarise, the final transfer matrix thus keeps enough information,
both about the mutual colouring of the sites and about the
connectivity of the clusters, to give the correct Boltzmann weights to
the different configurations, even for non-integer $Q$, and to follow
the evolution of the boundary of a particular set of clusters. These
boundaries are precisely the domain walls (see Fig.~\ref{DW2types}).

\section{Numerical results}

We have numerically diagonalised the transfer matrix in the DW
representation for cylinders and strips of width up to $L=11$
spins. We verified that the leading eigenvalue $\Lambda_0$ in the
ground state sector coincides with that of the FK transfer matrix,
including for non-integer $Q$.  As to the excitations
$\Lambda_{\ell_1,\ell_2}$, we explored systematically all possible
colouring combinations for up to $4$ marked spin clusters, for a
variety of values of the parameter $\kappa$, and for several different
boundary conditions (periodic, free, fixed, and mixed).

Finite-size approximations of the critical exponents $h(L)$ and
$\widetilde{h}(L)$ were extracted from the leading eigenvalue in each
sector, using standard CFT results
\cite{CardyBloteNightingale,JJreview}, and fitting both for the
universal corrections in $L^{-2}$ and the non-universal $L^{-4}$ term.
These approximations were further extrapolated to the $L \to \infty$
limit by fitting them to first and second order polynomials in
$L^{-1}$, gradually excluding data points corresponding to the
smallest $L$.  Error bars were obtained by carefully comparing the
consistency of the various extrapolations.

\begin{table}
\begin{center}
\begin{tabular}{r|rrrrrr}
  $p=\frac{\kappa}{4-\kappa}$ &  $(2,0)$ &  $(0,2)$ &  $(3,0)$ &  $(2,1)$ &  $(0,3)$ & $(0,1)^\ast$ \\ \hline
  $2$ &  2.01(1) &  5.99(2) &  2.97(4) &          &  8.94(2) & 1.000(0) \\ 
  $3$ &  4.01(1) &  7.99(2) &  6.02(2) &  8.04(2) & 11.98(3) & 1.500(1) \\
  $4$ &  5.93(2) & 10.01(2) &  8.89(5) & 10.93(5) & 15.05(4) & 1.992(2) \\
  $5$ &  7.77(4) & 12.09(8) & 11.6 (1) & 13.8 (1) & 18.2 (2) & 2.47 (2) \\[0.1cm]
Exact & $2(p-1)$ & $2(p+1)$ & $3(p-1)$ & $3p-1$   & $3(p+1)$ & $p/2$ \\
\end{tabular}
\end{center}
\caption{\label{tab1}
  Bulk critical exponents corresponding to five different DW configurations
  $(\ell_1,\ell_2)$, as functions of the parameter
  $p = \frac{\kappa}{4-\kappa}$, along with the conjectured exact expression
  (\ref{mainconj}). The last column labelled $(0,1)^\ast$ is for a single DW
  which is allowed to wrap around the periodic direction. The table entries
  give the value of $|\rho|$, when (\ref{Kac}) is rewritten as
  $h_{r,s} = (\rho^2-1)/(4p(p+1))$, with error bars shown in parentheses.} 
\end{table}

Representative final results for the bulk case (periodic boundary
conditions) are shown in Table~\ref{tab1}. Recall that
(\ref{mainconj}) is only valid for $(\ell_1,\ell_2)=(0,1)$ if the spin
cluster is forbidden to wrap around the periodic direction. The sector
where such wrapping is allowed is denoted $(0,1)^\ast$ in
Table~\ref{tab1}, and the corresponding critical exponent turns out to
be $h_{0,1/2}$.

\begin{table}
\begin{center}
\begin{tabular}{r|rrrr}
  $p=\frac{\kappa}{4-\kappa}$ &  $(2,0)$ &  $(1,2)$ &  $(2,1)$ &  $(3,0)$ \\ \hline
  $3$ &   7.00(3) & 19.0(2) & 15.16(4) & 11.13(4) \\
  $4$ &  10.89(7) & 24.8(2) & 20.9 (1) & 16.74(5) \\
  $5$ &  14.5 (1) & 31.1(2) & 26.7 (2) & 22.2 (3) \\[0.1cm]
Exact & $4p-5$    & $6p+1$  & $6p-3$   & $6p-7$   \\
\end{tabular}
\end{center}
\caption{\label{tab2}
  Boundary critical exponents with free boundary conditions, corresponding to
  four different DW configurations $(\ell_1,\ell_2)$, as functions of the
  parameter $p = \frac{\kappa}{4-\kappa}$, along with the conjectured exact
  expression (\ref{mainconj}). The table entries
  give the value of $|\rho|$, when (\ref{Kac}) is rewritten as
  $h_{r,s} = (\rho^2-1)/(4p(p+1))$, with error bars shown in parentheses.} 
\end{table}

Final results for the boundary case with free boundary conditions are
shown in Table~\ref{tab2}. The convention for the labels $(\ell_1,\ell_2)$
is the one stated below Eq.~(\ref{magn_exp}).

\section{Mixed boundary conditions}
\label{sec:mixed}

We recall that mixed boundary conditions have been defined by imposing
that the spins on the left boundary of the strip belong to a subset of
$Q_1$ different colours, with $0 < Q_1 \le Q$, while the spins on the
right boundary remain free.  In particular, $Q_1 = 1$ (resp.\ $Q_1 =
Q$) describes fixed (resp.\ free) boundary conditions. As above, the
convention for the labels $(\ell_1,\ell_2)$ is the one stated below
Eq.~(\ref{magn_exp}).

\begin{figure}[htbp]
 \begin{center}
  \includegraphics[width=0.49\textwidth]{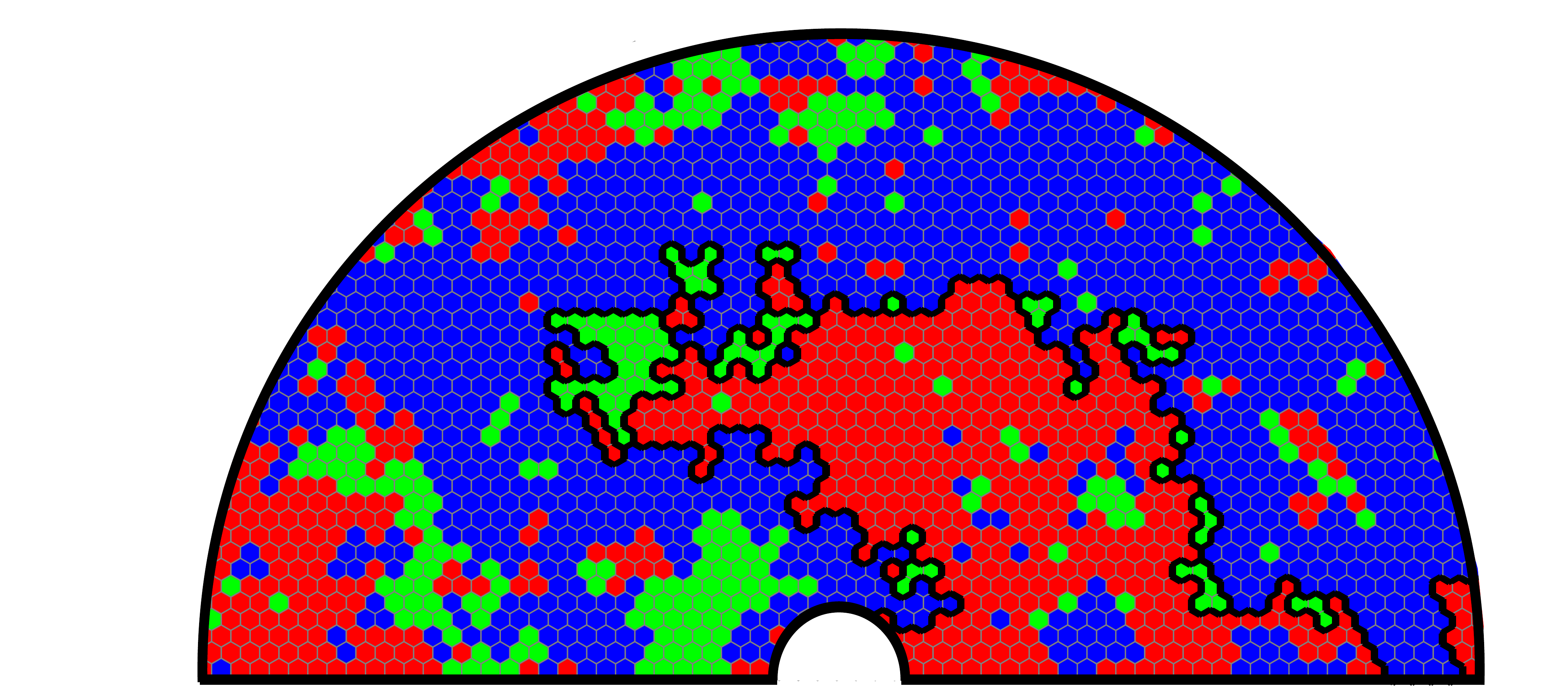}
  \includegraphics[width=0.49\textwidth]{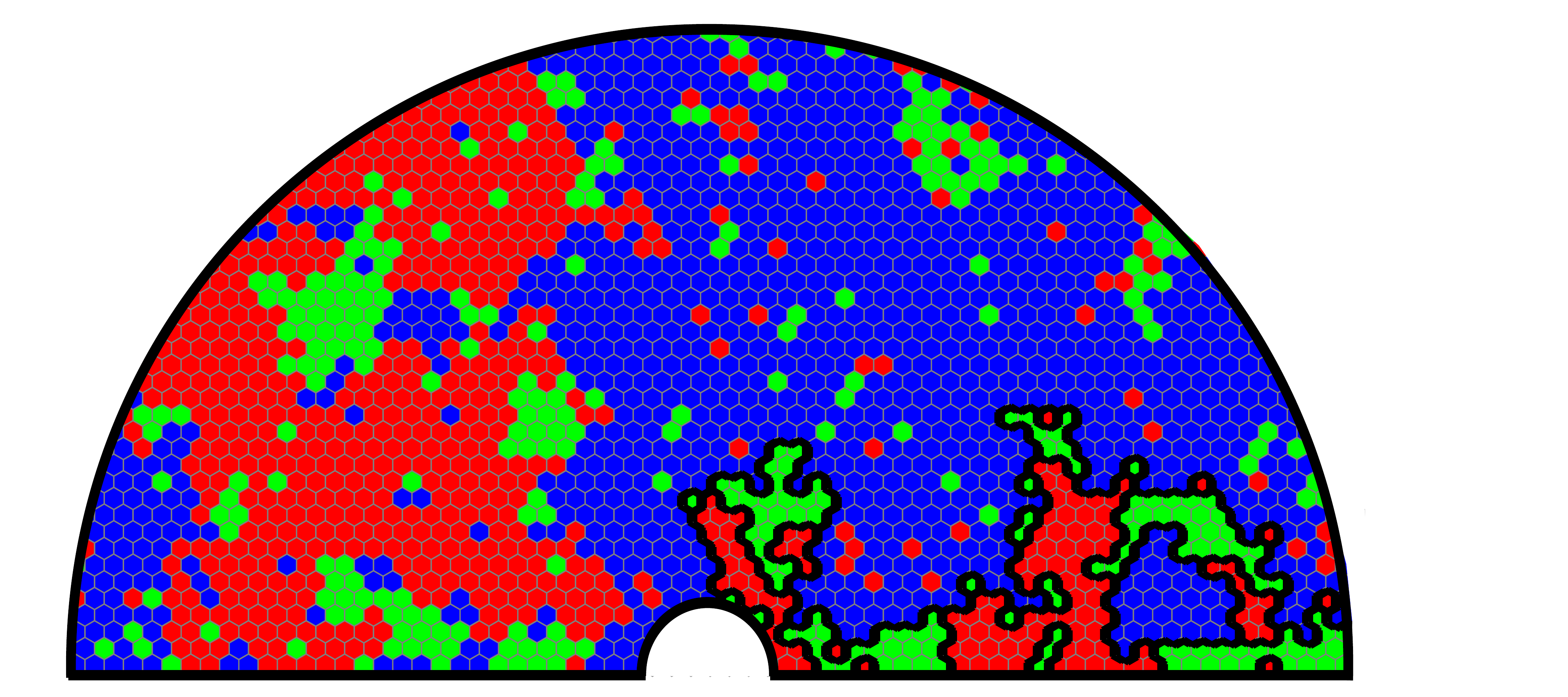}
 \end{center}
 \caption{Mixed boundary conditions in the half plane geometry for
   $Q=3$ and $Q_1 = 2$. The left (resp.\ right) panel illustrates the
   case $\epsilon = 1$ (resp.\ $\epsilon = -1$) where the propagating
   cluster, shown in blue colour, is allowed to (resp.\ not allowed
   to) touch the positive real axis.}
 \label{halfplane34}
\end{figure}

An important point is that the fixed colour of the leftmost marked
spin cluster may or may not be one of those allowed on the left
boundary.  For each choice of colours of the marked clusters, we
define a sign $\epsilon = 1$ (resp.\ $\epsilon = -1$) if the colour of
the leftmost cluster is $k_1 \le Q_1$ (resp.\ $k_1 > Q_1$), i.e., if
the leftmost cluster is allowed to (resp.\ not allowed to) touch the
left boundary. This sign closely parallels a construction in
\cite{CBL}, where the case $\epsilon=1$ (resp.\ $\epsilon = -1$) was
referred to as the blobbed (resp.\ unblobbed) sector. For free
boundary conditions, only the choice $\epsilon = 1$ is meaningful (at
least for $0 < Q_1 \le Q$, although it might be possible to give a
meaningful definition of the model outside this range, by a suitable
analytical continuation).
The issue of the sign $\epsilon = \pm 1$ is illustrated in
Fig.~\ref{halfplane34}.

One would expect the mixed boundary conditions to be given by the
conformal fusion of the operator $\Phi_{1+2(\ell_1-\ell_2),1+4
  \ell_1}$ that inserts the required number of DW, and the
free-to-mixed boundary condition changing operator $\Phi_{r_0,s_0}$.
The latter operator is responsible for the shift from $Q$ to $Q_1$
allowed colours on the boundary. It has been worked out in \cite{CBL}
in the context of a different set of geometrical observables (TL
loops), but since it should be representation-independent we can
take it over here. The dominant exponents for mixed boundary
conditions therefore follow from the standard CFT fusion rules
\cite{JJreview} as
\begin{equation}
 h_{r_0 + 2\epsilon(L_1-L_2),s_0 + 4\epsilon L_1} \,,
 \label{main_mixed}
\end{equation}
where $h_{r_0,s_0}$ is the dimension of the free-to-mixed
boundary condition changing operator $\Phi_{r_0,s_0}$.

The values of $(r_0,s_0)$ follow from \cite{CBL}, e.g., $(-1,-2)$ for
$Q_1=1$ and $(-\frac12,-1)$ for $Q_1=2$. We have verified these and
other cases by explicit numerical calculations. In particular, the
possible ambiguity of $\epsilon$ versus $-\epsilon$ in
(\ref{main_mixed}) is ruled out by the numerical results.
Moreover (\ref{main_mixed}) agrees with (\ref{mainconj}) for
free boundary conditions ($\epsilon=1$) as it should.


\section{Massless scattering description}

While it is far from obvious to derive (\ref{mainconj}) by CG methods,
minor progress can be achieved using a rather different set of
ideas. We restrict the discussion in this section to the bulk properties.

We can learn about the dynamics of DW in the critical theory by using
known information about the low-temperature ($K>K_{\rm c}$) phase of
the Potts model. Albeit non-integrable on the lattice, the
corresponding deformation by the operator $\Phi_{21}$ is integrable in
the continuum \cite{ChimZamo}. It can be described using a basic set
of kinks $K_{ab}$ separating two vacua, i.e., ordered regions where
the dominant value of the spin is $a$, resp.\ $b$. These kinks scatter
with a known S-matrix related to the BWM algebra \cite{ReadFendley}.
Importantly, the dynamics conserves the number of kinks: the process
$K_{ab}K_{bc} \to K_{ac}$ is forbidden (as in any elastic relativistic
scattering theory), although kinks do appear as bound states in
kink-kink processes.

Many properties of these kinks can be calculated using integrability
techniques. When the mass $m \to 0$ (i.e., $K\to K_c$), the S-matrix
provides a ``massless scattering'' description \cite{FenSal} of some
of the degrees of freedom of the critical theory itself. It is not
entirely clear what a kink, which is well-defined for $K \gg K_{\rm
  c}$, becomes at $K_{\rm c}$, but it is natural to expect that thick
DW are described by the propagation of two (or more) kinks such as
$K_{ab}K_{ba}$. As for thin DW, the potential existence of regions
where they are reduced to a single edge---that is a single
kink---suggest they have to do with more complicated processes
involving two kinks merging into one. We will not discuss them further
here.

It is an easy exercise to obtain the scaling dimension of thick DW
using the massless scattering description. Indeed, the fact that the
S-matrix satisfies relations from the BWM algebra allows us to
reexpress it in terms of the $a_2^{(2)}$ or Bullough-Dodd S-matrix
\cite{Efth}, for which the thermodynamic Bethe ansatz was studied in
\cite{SalWehe}. This describes as well the dynamics of the field
theory
\begin{equation}
 S = \frac{1}{8\pi} \int {\rm d}^2x \,
 \left[(\partial_x\Phi)^2+(\partial_y\Phi)^2 +
 g(2 {\rm e}^{-\frac{\rm i}{\sqrt{2}} \beta\Phi} + 
 {\rm e}^{{\rm i}\sqrt{2}\beta\Phi})\right]
\end{equation}
with $\beta^2 = \frac{\kappa}{4}$. Giving each kink the fugacity $Q-1$
produces the correct central charge
\begin{equation}
 c=1-\frac{3}{2} \frac{(\kappa-4)^2}{\kappa} \,,
\end{equation}
where each kink has a $U(1)$ charge equal to $0,\pm 1$. The scaling of
the sector with charge $j$ produces a gap $\Delta_j = \frac{j^2}{4
  \kappa}$, so the leading dimension is $\Delta_j-(1-c)/24 =
h_{j/2,0}$.  This agrees with $h_{-\ell_2,0}$ with $\ell_2=j/2$, so
there are two kinks per thick DW.

This argument validates (\ref{mainconj}) for $\ell_1 = 0$.

\section{Discussion}

The results (\ref{mainconj}) and (\ref{magn_exp}) can be used to
predict the fractal dimension of various geometrical objects
related to spin clusters. It should be possible to observe these
dimensions in Monte Carlo simulations, and possible in real experiments.

\subsection{Fractal dimensions}

The dimension of a spin cluster follows from (\ref{magn_exp}) as
\begin{eqnarray}
 d &=& {\rm min} (2,\,2-2h_{0,1/2}) \nonumber \\
   &=& \left \lbrace
   \begin{array}{ll}
   2 & \mbox{for } \kappa \in [2,\frac{8}{3}] \\
   \frac{(8+\kappa)(8+3\kappa)}{32\kappa}
     & \mbox{for } \kappa \in [\frac{8}{3},4] \\
   \end{array}
   \right. \label{res_d}
\end{eqnarray}
in agreement with \cite{DuplantierSaleur,Vanderzande}.

The boundary of a spin cluster has dimension
\begin{equation}
 d_{\rm b}=2-2h_{-1,0}=1+\frac{\kappa}{8} \,.
 \label{res_db}
\end{equation}
This is an example of a duality relation: $d_{\rm b}$ for a spin
cluster with SLE parameter $\kappa$ equals the dimension
$d_{\rm b}^{\rm FK}$ of the boundary of an FK
cluster at the dual parameter $\kappa^\ast = \frac{16}{\kappa}$. It is
moreover known that $d_{\rm b}^{\rm FK}$ at
parameter $\kappa$ coincides with the dimension $d_{\rm b}^{\rm EP}$ of
the FK cluster's
external perimeter (with fjords filled in) at parameter $\kappa^\ast$
\cite{Duplantierduality}. Combining these, (\ref{res_db}) means that
$d_{\rm b} = d_{\rm b}^{\rm EP}$, i.e., the dimension of the boundary of
a spin cluster equals that of the external perimeter of an FK cluster,
at the same $\kappa$. This
latter fact has very recently been verified numerically by Monte Carlo
simulations, including for non-integer values of $Q$
\cite{Zatelepin_Shchur}.

In the half-plane geometry, the intersection of the spin cluster containing
the origin with the real axis has the dimension
\begin{equation}
 d_{\rm s} = {\rm min} (1,\,1-h_{3,5}) =
 {\rm min} \left(1,\,8 - \frac{8}{\kappa} - \frac{3 \kappa}{2}\right) \,,
 \label{res_ds}
\end{equation}
as follows from (\ref{mainconj}).

\subsection{Thin and thick domain walls}

We have initially attached the epithets `thin' and `thick' to the two
types of domain walls based on considerations in the microscopic
model. Indeed, thin DW can narrow down to a single lattice edge on
the dual lattice, separating spins of different colour. Since
(\ref{mainconj}) implies that the two types of DW scale differently in
the continuum limit, one would expect that also the continuum
geometrical objects can somehow be characterised as `thin' and
`thick'. We now show that this is indeed the case.

According to (\ref{mainconj}), the set of points where a thin DW has
minimal width---i.e., one lattice spacing in the microscopic
formulation, or the two clusters separated by the DW `come close' in
the continuum limit in the sense of the small neighbourhoods defined
previously---has dimension
\begin{equation}
 d_1 = 2-2h_{2,4} = \frac{3}{8 \kappa} (4-\kappa) (5 \kappa-4) \,.
\end{equation}
This is analogous to the dimension of so-called `red' or `pivotal' bonds
in the theory of percolation and of FK clusters. The corresponding
result for a thick DW reads
\begin{equation}
 d_2 = {\rm max}(0,\, 2-2h_{-2,0}) = 0 \,.
\end{equation}
The fact that $d_1 \ge 0$ means that thin DW are indeed thin, in the
sense that they have a macroscopic number of loci of zero thickness in
the continuum limit. Likewise, the fact that $d_2 = 0$ means that
the loci where thick DW have zero thickness is a set of measure zero.

To summarise, there is therefore a consistency between the distinction of
the two types of DW microscopically and in the continuum limit, and this
ensures that they can (and do) scale differently.

One can also obtain dimensions analogous to $d_1$ and $d_2$ at the boundary.
Indeed, define $\widetilde{d}_1$ (resp.\ $\widetilde{d}_2$) to be the
fractal dimension of the set which is the intersection of the real axis
with the set of points where a thin (resp.\ thick) DW has minimal width.
In other words, these are the dimensions of pivotal bonds on the boundary.
By setting $\ell_1 = 2$ and $\ell_2 = 0$ (resp.\ $\ell_1 = 0$ and $\ell_2 = 2$)
in (\ref{mainconj}) we have then
\begin{eqnarray}
 \widetilde{d}_1 &=& {\rm min}(1,\, {\rm max}(0,\, 1 - h_{5,9})) \nonumber \\
                 &=& \left \lbrace
 \begin{array}{ll}
  1 & \mbox{for } \kappa \in [2,\frac{12}{5}] \\
  \frac{1}{\kappa} (3-\kappa)(5\kappa-8) & \mbox{for } \kappa \in [\frac{12}{5},3] \\
  0 & \mbox{for } \kappa \in [3,4] \\
 \end{array}
 \right.
\end{eqnarray}
respectively
\begin{equation}
 \widetilde{d}_2 = {\rm max}(0,\, 1-h_{-3,1}) = 0 \,.
\end{equation}

\subsection{The limit $Q=4$}

In the limit $Q \to 4$ (or $\kappa \to 4$), we have $d_1 \to 0$ and
$d_2 \to 0$. This means that the distinction between thin and thick DW
disappears in that limit. This is of course consistent with the fact
that $\kappa=4$ is a fixed point of the duality transformation
$\kappa^* = \frac{16}{\kappa}$, and so spin clusters and FK clusters
have identical properties.

A further manifestation of the indistinguishability of thin and thick
DW is that the bulk exponent (\ref{mainconj}) becomes
\begin{equation}
 h_{\ell_1-\ell_2,2\ell_1} = \frac{(\ell_1+\ell_2)^2}{4} \,.
\end{equation}
Indeed, this formula is invariant under the permutation of $\ell_1$
and $\ell_2$. Moreover, the exponent for $\ell$ FK clusters reads
$\frac{\ell^2}{4}$ for $Q=4$ \cite{JJreview,Nienhuis}, and so a spin
DW (whether thin or thick) behaves as a FK cluster, as expected.

\subsection{The case $Q=2$}

For the Ising model $Q=2$ (or $\kappa=3$) the absense of branchings
means that one thick DW equals two thin DW. Indeed the bulk exponent
becomes (\ref{mainconj}) becomes
\begin{equation}
 h_{\ell_1-\ell_2,2\ell_1} = \frac{4 \ell^2-1}{48}
\end{equation}
with $\ell = \ell_1 + 2 \ell_2$ in that case, and this agrees with the
exponent $h_{\ell/2,0}$ for $\ell$ loop strands \cite{JJreview,
  Nienhuis} in the dilute O($1$) model.

\subsection{The limit $Q \to 1$}

The limit $Q \to 1$ (or $\kappa \to \frac{8}{3}$) is often studied in
the context of FK clusters, since these become then bond percolation
clusters. Setting $Q=1$ for spin clusters has a more trivial meaning:
all the spins are simply in the same state, $\sigma_i = 1$. Therefore
we should have $d=2$ and $d_{\rm s}$ for $Q \to 1$. This is indeed
satisfied by (\ref{res_d}) and (\ref{res_ds}).

Apart from that, the $Q \to 1$ limit for spin clusters is more subtle.
One can tackle it by considering configurations of bond percolation
on the square lattice where one picks up {\em one} single percolation
cluster. The spins on the vertices sitting on this cluster are given
a certain colour. All the other spins in the system are given another
single colour. Then the boundary of the coloured {\em spin} cluster
is a single loop with fractal dimension $d_f = 4/3$, as expected from
the duality $\kappa^* = 16/\kappa$. It is, however, remarkable that
we can find such a simple construction of the {\em spin} cluster
for $Q=1$ on the lattice. For more interfaces one has to pick up
more percolation clusters, and give them different colours.

Finally, we note that for $Q=1$ we have
\begin{equation}
 h_{\ell_1-\ell_2,2\ell_1} = \frac{\ell^2-1}{24} \,,
\end{equation}
which agrees with $h_{0,\ell/2}$ if one sets $\ell=\ell_1+3\ell_2$.
This seems to indicate that for percolation, 
a thick DW equals {\em three} thin DWs (or TL loop strands)---a curious
result for which we have no convincing explanation at present.

\section{Conclusion}

We have defined a set of geometrical observables based on the
branching domain walls of the Potts model for any real $0 \leq Q \leq
4$. These observables are defined starting from the DW expansion
(\ref{PottsDW}) of the partition function.  We have studied
numerically these objects for non-integer $Q$ using a transfer matrix
formulation. Our results are compatible with conformal invariance of
these observables and we have given a set of conjectures for the
corresponding bulk and boundary critical exponents.

\section*{Acknowledgments}

We thank G.\ Delfino and P.\ Fendley for discussions. This work was
supported by the Agence Nationale de la Recherche (grant
ANR-06-BLAN-0124-03).


\section*{References}

\begin{thebibliography}{99}

\bibitem{Wu08} F.Y.\ Wu, {\em Exactly solved models: A journey in
    statistical mechanics} (World Scientific, Singapore, 2008).

\bibitem{Baxter82} R.J.\ Baxter, {\em Exactly solved models in
    statistical mechanics} (Academic Press, London, 1982).

\bibitem{Eisenriegler} E. Eisenriegler, {\em Polymers near Surfaces} (World
  Scientific, Singapore, 1993).

\bibitem{FK72} C.M.\ Fortuin and P.W.\ Kasteleyn,
  Physica {\bf 57}, 536 (1972).

\bibitem{JJreview} J.L.\ Jacobsen,
  Lect.\ Notes Phys.\ {\bf 775}, 347--424 (2009).

\bibitem{BBreview} M.\ Bauer and D.\ Bernard,
  Phys.\ Rep.\ {\bf 432}, 115 (2006).

\bibitem{Nienhuis} B.\ Nienhuis, \textit{Loop Models}, Les Houches Summer School: Volume 89, July 2008. (Oxford University Press, 2010)

\bibitem{backbone} Y.\ Deng, H.W.J.\ Bl\"ote and B.\ Nienhuis,
  Phys.\ Rev.\ E {\bf 69}, 026114 (2004).

\bibitem{shortest_path} Y.\ Deng, W.\ Zhang, T.M.\ Garoni,
  A.D.\ Sokal and A.\ Sportiello,
  Phys.\ Rev.\ E {\bf 81}, 020102(R) (2010).

\bibitem{DuplantierSaleur}  B.\ Duplantier and H.\ Saleur,
  Phys.\ Rev.\ Lett. {\bf 63}, 2536 (1989).
 
\bibitem{Vanderzande} C.\ Vanderzande,
  J.\ Phys.\ A {\bf 25}, L75 (1992). 

\bibitem{Coniglio} A.\ Coniglio and F.\ Peruggi,
  J.\ Phys.\ A {\bf 15}, 1873 (1982). 

\bibitem{Qian} X.\ Qian, Y.\ Deng and H.W.J.\ Bl\"ote,
  Phys.\ Rev.\ B {\bf 71}, 144303 (2005).

\bibitem{Janke} W.\ Janke and A.\ Schakel,
 Braz.\ J.\ Phys.\ {\bf 36}, 708 (2006);
 Phys.\ Rev.\ E {\bf 71}, 036703 (2005);
 Nucl.\ Phys.\ B {\bf 700}, 385 (2004).

\bibitem{Stella} A.L.\ Stella and C.\ Vanderzande,
  Phys.\ Rev.\ Lett.\ {\bf 62}, 1067 (1989);
%
  J.\ Phys.\ A {\bf 22}, L445 (1989).

\bibitem{Duplantierduality} B. Duplantier, 
 Phys.\ Rev.\ Lett.\ {\bf 84}, 1363 (2000).
 
\bibitem{CardyGamsa} A.\ Gamsa and J.\ Cardy,
 J.\ Stat.\ Mech.\ P08020 (2007).

\bibitem{FendleyModels} P. Fendley,
 Annals of Physics {\bf 323}, 3113 (2008).

\bibitem{Picco_Santachiara} M.\ Picco, R.\ Santachiara and A.\ Sicilia,
 J.\ Stat.\ Mech.\ (2009) P04013;
 M.\ Picco and R.\ Santachiara, arXiv:1005.0493.

\bibitem{Fendley} P.\ Fendley,
 J.\ Phys.\ A {\bf 39}, 15445 (2006).

\bibitem{ReadFendley} P.\ Fendley and N.\ Read,
 J.\ Phys.\ A {\bf 35}, 10675 (2002).

\bibitem{Letter} J.\ Dubail, J.L.\ Jacobsen and H.\ Saleur,
  arXiv:1008.1216.

\bibitem{CBL} J.L.\ Jacobsen and H.\ Saleur,
  Nucl.\ Phys.\ B {\bf 788}, 137 (2008);
  J.\ Dubail, J.L.\ Jacobsen and H.\ Saleur,
  Nucl.\ Phys.\ B 813, 430 (2009); {\em ibid.} {\bf 827}, 457 (2010).

\bibitem{Sweeny} M.\ Sweeny,
  Phys.\ Rev.\ B {\bf 27}, 4445 (1983).

\bibitem{Chayes_Machta} L.\ Chayes and J.\ Machta,
  Physica A {\bf 254}, 477 (1998).

\bibitem{Deng_Blote} Y.\ Deng, X.\ Qian and H.W.J.\ Bl\"ote,
  Phys.\ Rev.\ E {\bf 80}, 036707 (2009).

\bibitem{Zatelepin_Shchur} A.\ Zatelepin and L.\ Shchur,
  arXiv:1008.3573.

\bibitem{CardyBloteNightingale} H.W.J.\ Bl\"ote,
  J.L.\ Cardy and M.P.\ Nightingale,
  Phys.\ Rev.\ Lett.\ {\bf 56}, 742 (1986). 

\bibitem{Blote82} H.W.J.\ Bl\"ote and M.P.\ Nightingale,
 Physica A {\bf 112}, 405 (1982).

\bibitem{ChimZamo} L.\ Chim and A.\ Zamolodchikov,
 Int.\ J.\ Mod.\ Phys.\ A {\bf 7}, 5317 (1992).

\bibitem{FenSal} P.\ Fendley and H.\ Saleur, in Gava {\em et al.} (eds.), 
 Proceedings of the Trieste Summer School in High Energy Physics and Cosmology
 (World Scientific, 1993).

\bibitem{Efth} 
 C.J.\ Efthimiou,
 Nucl.\ Phys.\ B {\bf 398}, 697 (1993).

\bibitem{SalWehe} H.\ Saleur and B.\ Wehefritz-Kaufmann,
 Nucl.\ Phys.\ B {\bf 628}, 407 (2002).


 
\end{thebibliography}
\end{document}